\documentclass[pdftex,twocolumn,epjc3]{svjour3}

\RequirePackage[T1]{fontenc}

\usepackage{graphicx}
\usepackage{subfigure}
\RequirePackage{mathptmx}
\RequirePackage{flushend}
\RequirePackage[numbers,sort&compress]{natbib}
\usepackage[range-units=single,range-phrase=--,uncertainty-mode=separate,print-unity-mantissa=false]{siunitx}
\usepackage{xspace}
\usepackage{multirow}
\usepackage{lineno}
\usepackage{microtype}
\usepackage[colorlinks,allcolors=blue]{hyperref}
\journalname{Eur. Phys. J. C}

\newcommand{\gam}{$\gamma$\xspace}
\newcommand{\gp}{\textsc{Gammapy}\xspace}

\newcommand{\rxj}{RX~J1713.7$-$3946\xspace}
\newcommand{\velax}{Vela~X\xspace}
\newcommand{\wld}{Westerlund~1\xspace}
\newcommand{\hessjn}{HESS~J1908+063\xspace}
\newcommand{\ehwc}{eHWC~J1907+063\xspace}

\begin{document}

\title{Prospects for combined analyses of hadronic emission from \gam-ray sources in the Milky Way with CTA and KM3NeT}

\author{
T.~Unbehaun\thanksref{ECAP,emailtim} \and
L.~Mohrmann\thanksref{MPIK,emaillars} \and
S.~Funk\thanksref{ECAP}\\(authors of the CTA Consortium)
\\and\\
S.~Aiello\thanksref{a} \and
A.~Albert\thanksref{b,be} \and
S.~Alves Garre\thanksref{c} \and
Z.~Aly\thanksref{d} \and
A.~Ambrosone\thanksref{f,e} \and
F.~Ameli\thanksref{g} \and
M.~Andre\thanksref{h}
% ...
\\(KM3NeT Collaboration\thanksref{emailkm3})
}

\thankstext{emailtim}{e-mail: \href{mailto:tim.unbehaun@fau.de}{tim.unbehaun@fau.de}}
\thankstext{emaillars}{e-mail: \href{mailto:lars.mohrmann@mpi-hd.mpg.de}{lars.mohrmann@mpi-hd.mpg.de}}
\thankstext{emailkm3}{e-mail: \href{mailto:km3net-pc@km3net.de}{km3net-pc@km3net.de}}

\institute{
Erlangen Centre for Astroparticle Physics, University of Erlangen-Nuremberg, Nikolaus-Fiebiger-Str.~2, 91058 Erlangen, Germany \label{ECAP} \and
Max-Planck-Institut f\"ur Kernphysik, Saupfercheckweg~1, 69117~Heidelberg, Germany \label{MPIK} \and
INFN, Sezione di Catania, Via Santa Sofia 64, Catania, 95123 Italy \label{a} \and
Universit{\'e}~de~Strasbourg,~CNRS,~IPHC~UMR~7178,~F-67000~Strasbourg,~France \label{b} \and
IFIC - Instituto de F{\'\i}sica Corpuscular (CSIC - Universitat de Val{\`e}ncia), c/Catedr{\'a}tico Jos{\'e} Beltr{\'a}n, 2, 46980 Paterna, Valencia, Spain \label{c} \and
Aix~Marseille~Univ,~CNRS/IN2P3,~CPPM,~Marseille,~France \label{d} \and
INFN, Sezione di Napoli, Complesso Universitario di Monte S. Angelo, Via Cintia ed. G, Napoli, 80126 Italy \label{e} \and
Universit{\`a} di Napoli ``Federico II'', Dip. Scienze Fisiche ``E. Pancini'', Complesso Universitario di Monte S. Angelo, Via Cintia ed. G, Napoli, 80126 Italy \label{f} \and
INFN, Sezione di Roma, Piazzale Aldo Moro 2, Roma, 00185 Italy \label{g} \and
Universitat Polit{\`e}cnica de Catalunya, Laboratori d'Aplicacions Bioac{\'u}stiques, Centre Tecnol{\`o}gic de Vilanova i la Geltr{\'u}, Avda. Rambla Exposici{\'o}, s/n, Vilanova i la Geltr{\'u}, 08800 Spain \label{h}
% ...
\and
Universit{\'e} de Haute Alsace, rue des Fr{\`e}res Lumi{\`e}re, 68093 Mulhouse Cedex, France \label{be}
% ...
}

\date{\today}

\onecolumn

\maketitle

\twocolumn    


\begin{abstract}
  The Cherenkov Telescope Array and the KM3NeT neutrino telescopes are major upcoming facilities in the fields of \gam-ray and neutrino astronomy, respectively.
  Possible simultaneous production of \gam rays and neutrinos in astrophysical accelerators of cosmic-ray nuclei motivates a combination of their data.
  We assess the potential of a combined analysis of CTA and KM3NeT data to determine the contribution of hadronic emission processes in known Galactic \gam-ray emitters, comparing this result to the cases of two separate analyses.
  In doing so, we demonstrate the capability of \gp, an open-source software package for the analysis of \gam-ray data, to also process data from neutrino telescopes.
  For a selection of prototypical \gam-ray sources within our Galaxy, we obtain models for primary proton and electron spectra in the hadronic and leptonic emission scenario, respectively, by fitting published \gam-ray spectra.
  Using these models and instrument response functions for both detectors, we employ the \gp package to generate pseudo data sets, where we assume 200 hours of CTA observations and 10 years of KM3NeT detector operation.
  We then apply a three-dimensional binned likelihood analysis to these data sets, separately for each instrument and jointly for both.
  We find that the largest benefit of the combined analysis lies in the possibility of a consistent modelling of the \gam-ray and neutrino emission.
  Assuming a purely leptonic scenario as input, we obtain, for the most favourable source, an average expected 68\% credible interval that constrains the contribution of hadronic processes to the observed \gam-ray emission to below 15\%.
\end{abstract}

\section{Introduction}
We live in the era of multi-messenger astrophysics \cite{Meszaros2019}.
Long anticipated, this paradigm advocates that unique insights about astrophysical objects and processes may be gained through the joint consideration of information carried by different messengers: photons, neutrinos, cosmic rays (CRs), and gravitational waves.
In the past years, it has truly come to fruition, yielding the first promising results \cite{LIGO2017,IceCube2018}.

Astrophysical objects in the Milky Way are not expected to produce gravitational waves detectable by current-generation instruments (but by next-generation detectors, see \cite{Gossan2022}).
Galactic CRs provide important energetic constraints on their source population, but cannot be used to directly study Galactic objects because they are deflected by magnetic fields.
Hence, one needs to resort to photons and neutrinos to employ multi-messenger astrophysics for the study of individual Galactic objects.

Indeed, besides its conceptual attractiveness, the combined study of very-high-energy (VHE; $E>\SI{100}{GeV}$) \gam rays and TeV--PeV neutrinos from Galactic sources is well motivated:
they are expected to be produced simultaneously in `hadronic accelerators', where accelerated CR nuclei interact with ambient gas producing pions (and other mesons) that subsequently decay into \gam rays and neutrinos.
In the following, this process is labelled `PD', for pion decay.
The situation is different in `leptonic accelerators', where Inverse Compton (IC) up-scattering of photons by CR electrons leads to VHE \gam-ray emission without neutrino production.
This implies that the detection of high-energy neutrinos coming from astrophysical objects is decisive in identifying them as hadronic accelerators \cite{Halzen2013}.
Nevertheless, observations of VHE \gam-ray emission from the same objects are indispensable, as they provide a much higher detection sensitivity and allow a measurement of the spectrum and morphology of the emission in greater detail.
This, in turn, enables a realistic estimation of the expected flux of neutrinos, and the possibility of detecting it with current (or planned) neutrino telescopes.
The latter exercise has been carried out by various authors in the past (see, e.g.\ \cite{Vissani2006,Kistler2006,Villante2008,Kappes2007,GonzalezGarcia2014,Halzen2017,Celli2017}).

In this work, we focus on the Cherenkov Telescope Array (CTA) \cite{CTA2018,Hofmann2023} and the KM3NeT neutrino telescopes \cite{KM3NeT_LoI_2016}, as major upcoming facilities for VHE \gam-ray and neutrino astronomy, respectively.
CTA will be built at La Palma, Spain, and Paranal, Chile, covering the Northern and Southern sky, respectively.
Our prime targets of interest being Galactic \gam-ray sources, which are more easily observed from the Southern hemisphere, we consider only the site in Chile (CTA-South) for our study.
At this site, the installation of $\sim$50 imaging atmospheric Cherenkov telescopes (IACTs) of two different sizes, covering the energy range between \SI{100}{GeV} and \SI{300}{TeV}, is foreseen.
IACTs detect \gam rays by measuring the faint flash of Cherenkov light that is emitted by secondary particles in the air shower that is launched when the primary \gam ray hits the atmosphere of the Earth.
Compared to current-generation arrays of IACTs, CTA is projected to provide a ten-fold increase in sensitivity.

KM3NeT is a research infrastructure in the Mediterranean, consisting of neutrino telescopes installed in the deep sea at different locations.
The `Oscillation Research with Cosmics in the Abyss' (ORCA) detector, with its dense instrumentation, will focus on the study of neutrino properties measuring atmospheric neutrinos \cite{KM3NeT_ORCA_2022}.
Here, we only consider the `Astroparticle Research with Cosmics in the Abyss' (ARCA) detector, which targets the detection of high-energy astrophysical neutrinos with TeV-PeV energies.
Hereafter, we will use the term `KM3NeT' to refer to the ARCA telescope.
It is currently under construction off-shore Sicily, Italy, and will ultimately consist of two building blocks comprising 115 vertical detection units each.
Each detection unit carries 18 digital optical modules (DOMs) \cite{KM3NeT_DOM_2014,KM3NeT_DOM_2016,KM3NeT_DOM_2022} with a vertical spacing of \SI{36}{\meter} and is about \SI{700}{\meter} tall.
The units are arranged on a grid with about \SI{90}{\meter} spacing between them.
The DOMs contain light sensors that detect Cherenkov light radiated by secondary particles created in interactions of high-energy neutrinos in or near the detector.
Of particular interest for this work are muons created in charged-current interactions of muon neutrinos, as their long propagation distances of up to several km in the water allow a precise reconstruction of the direction of the incoming neutrino.
Compared to the largest existing neutrino telescope, the IceCube Neutrino Observatory \cite{Ahlers2017,Ahlers2018}, KM3NeT utilises water instead of ice as detector medium.
This reduces scattering of the Cherenkov light and is expected to lead to an improved angular resolution \cite{KM3NeT_LoI_2016}.
Its location in the Northern hemisphere implies that neutrinos potentially emitted by many Galactic \gam-ray sources would reach KM3NeT through the Earth, which is advantageous for the suppression of atmospheric muon background events \cite{KM3NeT_PSsens_2019}.
Therefore, the combination of CTA-South and KM3NeT data for the study of Galactic objects appears very natural.

The discovery potential for extended Galactic sources by KM3NeT, in relation to the constraining power of CTA, has been investigated in \cite{Ambrogi2018}.
In this paper, we demonstrate how a combined analysis of CTA and KM3NeT data can be used to constrain physical properties of \gam-ray sources in our Galaxy, with special attention to the contribution of hadronic emission processes.
To this end, using Monte Carlo simulations as input, we have prepared instrument response functions (IRFs) for the KM3NeT detector and stored them in the same `GADF' data format\footnote{see \url{https://gamma-astro-data-formats.readthedocs.io} and \cite{Deil2022}} used for the publicly available CTA IRFs.
Then we have employed the \gp\footnote{\url{https://gammapy.org}} package (version 0.17; \cite{Deil2017,Deil2020}) to generate pseudo data sets based on the obtained IRFs and to perform a joint 3D likelihood fit on these data sets.
Though \gp is still under development and application of the 3D likelihood method in IACT data analysis represents a recent approach, both have been validated using a public IACT data set \cite{Mohrmann2019}.

\begin{figure}
    \centering
    \includegraphics[width=0.99\columnwidth]{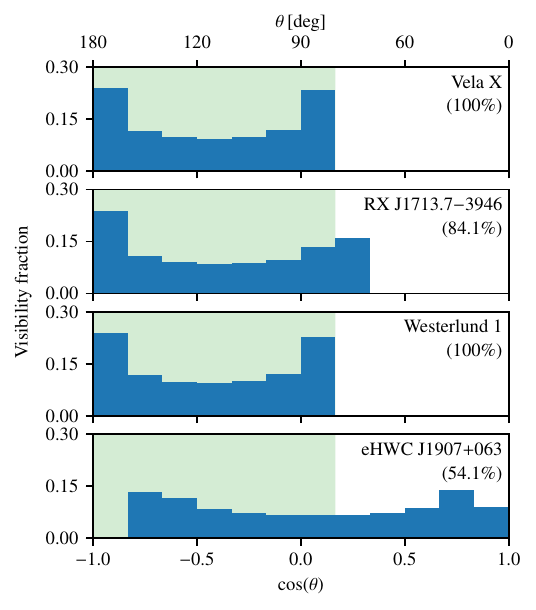}
    \caption{
      Source visibility with KM3NeT.
      Shown is the fraction of time that each source is visible under a zenith angle $\theta$, over the course of one year.
      The green-shaded area indicates the zenith angle range used in the analysis and the percentage value in parentheses specifies the fraction that each source is visible within this range.
    }
    \label{fig:visibility}
\end{figure}

We apply the analysis to a selection of prototypical sources that are promising candidates for the emission of high-energy neutrinos (see Table~\ref{tab:sources}).
We motivate our choice briefly in the following:
\begin{itemize}
\item \velax is a pulsar wind nebula bright in \gam rays \cite{HESS_VelaX}, associated with the well-known Vela pulsar.
Pulsars being copious producers of electrons and positrons, the \gam-ray emission from \velax is expected to be largely due to IC emission, and no associated neutrino emission is expected.
However, also mixed, lepto-hadronic models have been considered in the past (e.g.\ \cite{Horns2006,Zhang2009}), leaving room for some neutrino emission from this source.
\item \rxj is a shell-type supernova remnant that emits \gam rays in excess of \SI{10}{TeV} \cite{HESS_RXJ}.
The emission has been modelled according to leptonic, hadronic, and mixed scenarios (see e.g.\ \cite{Fukui2021,Cristofari2021} for recent studies).
The observation (but also non-observation) of neutrinos from \rxj could therefore yield important clues about acceleration processes at play.
\item \wld is the most massive young stellar cluster in the Milky Way \cite{Clark2005}, and is considered the most likely counterpart of the VHE \gam-ray source HESS~J1646$-$458 \cite{HESS_Wd1,HESS_Wd1_2022}.
Massive stellar clusters have recently been hypothesized as PeVatrons \cite{Aharonian2019}, making \wld a good candidate for high-energy neutrino emission.
\item \ehwc, also known as \hessjn, is an unidentified \gam-ray source \cite{HESS_J1908} that was recently detected above energies of \SI{100}{TeV} by the High Altitude Water Cherenkov Observatory (HAWC) \cite{HAWC2020} as well as by the Large High Altitude Air Shower Observatory\mbox{ } (LHAASO) \cite{LHAASO2021}.
Like in the case of \rxj, observations with neutrino telescopes could help to constrain the nature of the source.
\end{itemize}
Our selection of sources is neither a complete list of promising targets for the emission of neutrinos in our Galaxy, nor does it comprise only the most promising ones.
Rather, we have aimed for a selection of different types of \gam-ray sources that furthermore exhibit favourable locations for the observation with CTA-South and KM3NeT.
Figure~\ref{fig:visibility} shows the visibility of all sources for KM3NeT, as a function of the local zenith angle\footnote{
  The zenith angle $\theta$ refers to the location of the source.
  For $\theta=0^\circ$ the source is above the detector and produces vertically down-going neutrinos, while for $\theta=180^\circ$ the source is located on the opposite side of the Earth and produces vertically up-going neutrinos.
}.
For CTA, within one year, the sources are observable above an altitude angle of 50$^\circ$ for a maximum time of $\sim$\SI{400}{h} (\velax), $\sim$\SI{510}{h} (\rxj), $\sim$\SI{500}{h} (\wld), and $\sim$\SI{380}{h} (\ehwc).

\begin{table*}
  \caption{Galactic \gam-ray sources investigated in this work.}
  \label{tab:sources}
  \centering
  \begin{tabular}{lcccccc}
    \hline
    Designation & Type & Spatial model & $r$ & Declination & Distance & Reference\\
     & & & (deg)& (deg) & (kpc) & \\\hline
    \velax & PWN & disk & 0.8 & $-45.19$ & 0.29 & \cite{HESS_VelaX} \\
    \rxj & SNR & disk & 0.6 & $-39.69$ & 1 & \cite{HESS_RXJ} \\
    \wld & SC & disk & 1.1 & $-45.85$ & 3.9 & \cite{HESS_Wd1} \\
    \ehwc & UNID & Gaussian & 0.67 & $+06.18$ & 2.37 & \cite{HAWC2020} \\\hline
  \end{tabular}
  
  {\raggedright
    `Type' refers to the source type; PWN = pulsar wind nebula; SNR = supernova remnant; SC = stellar cluster; UNID = unidentified.
    `Spatial model' specifies which type of spatial model is used in the analysis (cf.\ section~\ref{sec:dataset_gen}).
    $r$ denotes the radius of the disk in case of a disk model and the width of the Gaussian in case of a Gaussian model.\par
  }
\end{table*}

Finally, we note that the IRFs we derived and used in this work are not representative of the final sensitivity of CTA and KM3NeT.
They are based on preliminary simulations and event selections that will likely be improved in the future.
In particular, the IRFs for KM3NeT are based on a point-source analysis as presented in \cite{Muller2021}.
This does not impact the conclusions drawn in this work, which are focused more on the conceptual benefits of a combined analysis rather than on numerical results.

The paper is structured as follows.
In Sect.~\ref{sec:methods}, we introduce our methodology: the computation of IRFs (Sect.~\ref{sec:irfs}), the preparation of input models (Sect.~\ref{sec:models}), the generation of pseudo data sets (Sect.~\ref{sec:dataset_gen}), the combined likelihood analysis (Sect.~\ref{sec:analysis}), and the derivation of constraints on hadronic contributions (Sect.~\ref{sec:computation-of-cl}).
The results of the analysis are then presented and discussed in Sect.~\ref{sec:results}, before we conclude the paper in Sect.~\ref{sec:conclusion}.

\section{Methodology}
\label{sec:methods}

\subsection{Instrument response functions}
\label{sec:irfs}
Given a physical source model, the IRFs for each experiment allow us to compute how this source would appear in the detector.
In our case, the relevant IRFs comprise the effective area, the energy dispersion, and the point spread function (PSF), reflecting the sensitivity, energy resolution, and angular resolution of the instrument, respectively.
Additionally, background templates that yield the expected number of mis-classified background events, arising from CR-induced atmospheric air showers, are necessary.

For IACTs, the IRFs are typically stored as a function of the true \gam-ray energy and of the true angular offset of the events with respect to the pointing direction of the telescopes (`offset angle').
The IRFs also depend on the angle with respect to zenith of the pointing position of the telescopes.
However, because the variation within a specific observation run (of typically \SI{30}{\minute} duration and a field of view of $\sim 5^\circ\times 5^\circ$) is small, IRFs for the average zenith angle of the observation run are commonly employed.
For CTA we use the publicly available `Prod 5' IRFs\footnote{See \url{https://www.cta-observatory.org/science/ctao-performance}.} for the southern array at 20$^\circ$ zenith angle and averaged over azimuth angle \cite{CTA2021}.
The IRFs for KM3NeT have been custom-generated for this study, as detailed in the following section.

\subsubsection{Generation of KM3NeT IRFs}
The KM3NeT IRFs are based on extensive simulations of neutrinos and anti-neutrinos\footnote{Hereafter, we will use the term `neutrinos' to refer to both neutrinos and anti-neutrinos, unless explicitly stated otherwise.} that interact in or near the detector.
We focus on charged-current interactions of muon neutrinos only, as they give rise to long-range muons that appear as characteristic, track-like events in the detector.
This leads to a good angular resolution ($<0.3^\circ$ for energies $>\SI{10}{TeV}$), which helps in suppressing background events (but see also \cite{Sudoh2023}).
The neutrino events have been simulated with the \textsc{gSeaGen} software \cite{KM3NeT_MCsoft_2020} based on the GENIE neutrino generator \cite{Andreopoulos2010}, which allows the simulation of interactions of all neutrino flavours in the media around the detector.

On the level of a single optical module, the decay of $^{40}\mathrm{K}$ as well as bioluminescence are relevant sources of noise.
Due to the design of the optical modules, which contain multiple photo-sensors each, these backgrounds can however be suppressed very efficiently by requiring a local coincidence between the photo-sensors \cite{KM3NeT_DOM_2016}.
On the analysis level, two types of background events are relevant for KM3NeT: neutrinos and muons, both created in CR-induced atmospheric air showers.
Both can be further classified as `conventional' -- resulting mostly from the decays of pions and kaons -- and `prompt' -- resulting from the decays of heavy hadrons and light vector mesons.
The former exhibit a steeper energy spectrum, because their parent particles have a non-negligible chance to re-interact with air molecules, rather than to decay.

To predict the rate of atmospheric neutrino events, we use the `HKKMS' model \cite{Honda2007} for conventional neutrinos and the `ERS' model \cite{Enberg2008} for prompt neutrinos.
Both models are based on outdated parametrisations of the primary CR flux and have been corrected as described in \cite{IceCube2014} to conform with the `H3a' parametrisation from \cite{Gaisser2012}.
We note that there is also the possibility of a CR composition around the `knee' feature in the CR spectrum that is heavier than predicted by the H3a model.
This scenario is discussed in more detail in \cite{Mascaretti2020}, but not investigated further here.

\begin{figure}
    \centering
    \includegraphics[width=0.99\columnwidth]{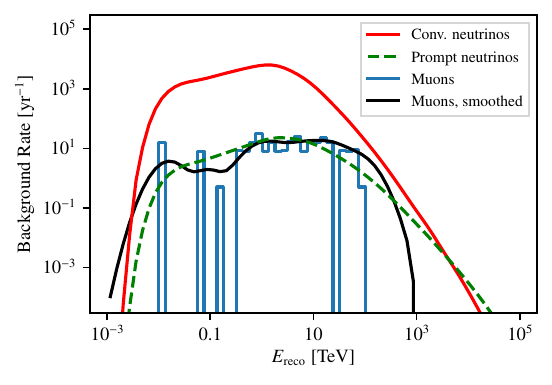}
    \caption{
      Background event rates in KM3NeT as a function of reconstructed energy $E_\mathrm{reco}$.
      The atmospheric neutrino rates are integrated over all zenith angles in the analysis region (i.e.\ $\theta_\mathrm{reco}>80^\circ$).
      The atmospheric muon rate is shown for the zenith angle bin $80^\circ-90^\circ$ only, since it is completely negligible for larger angles.
      The black line displays the smoothed curve used in the analysis.
    }
    \label{fig:bkg}
\end{figure}

The resulting event rates of conventional and prompt atmospheric neutrinos, integrated over relevant zenith angles, are shown in Fig.~\ref{fig:bkg}.
The background of atmospheric muons has been estimated using dedicated simulations of muons using the MUPAGE package \cite{Becherini2006,Carminati2008,ANTARES2021}.
While there is in principle also a potential background due to diffuse astrophysical neutrinos not connected to the studied source itself \cite{IceCube2022a}, this background can be safely neglected here.

All simulated events are reconstructed using a track reconstruction algorithm and subsequently undergo a selection procedure based on the reconstruction quality and a classification algorithm using boosted decision trees (BDTs), as detailed in \cite{Muller2021}.
In order to suppress the background of atmospheric muons, which always arrive from above the detector, we restrict the analysis region to reconstructed zenith angles $\theta_\mathrm{reco}>80^\circ$.
Thus, only very few atmospheric muon events remain in the final sample, almost all concentrated close to the horizon region ($80^\circ<\theta_\mathrm{reco}<90^\circ$), see the blue histogram in Fig.~\ref{fig:bkg}.
In order to avoid interpolation problems due to empty bins in the histogram, we fit a spline curve to the histogram and use this curve to predict the expected rate of atmospheric muon events (black line).
We note that due to insufficient simulation statistics, the exact shape of the distribution at energies below \SI{1}{TeV} should not be trusted.
Because the muon background is sub-dominant compared to the atmospheric neutrino background by several orders of magnitude at these energies, however, this does not affect our results.

The KM3NeT IRFs mainly depend on neutrino energy and zenith angle.
To be able to store the IRFs in the (IACT-centred) GADF data format, we utilise the offset-angle axis defined there to describe the dependence of the KM3NeT IRFs on the zenith angle.
The IRFs are then generated by creating histograms of the appropriate event properties (e.g.\ the angle between the reconstructed and true neutrino direction in case of the PSF) and applying corresponding normalisation factors.
For the effective area IRF, we use 48 logarithmic bins in true energy between \SI{100}{GeV} and \SI{100}{PeV} and 12 zenith angle bins linear in true $\cos(\theta)$.
The energy dispersion and PSF IRFs -- featuring one more dimension than the effective area -- are created with twice the bin size, in order to ensure sufficient statistics in each bin.
In the analysis, a linear interpolation between the individual bins is performed.
IRFs for neutrinos and anti-neutrinos are derived separately and subsequently averaged, assuming equipartition of the source flux between the two.

\subsubsection{Comparison of IRFs}

\begin{figure}
    \centering
    \includegraphics[width=0.99\columnwidth]{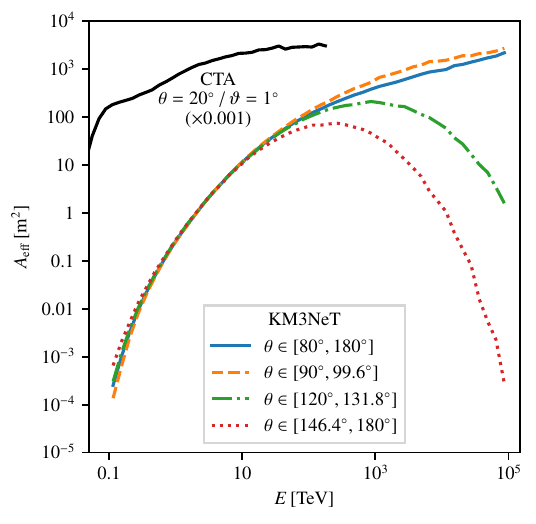}
    \caption{
      Comparison of effective areas.
      The CTA effective area is shown for a zenith angle of $\theta=20^\circ$ and an offset angle from the pointing direction of $\vartheta=1^\circ$.
      For KM3NeT, average effective areas for the full analysis region as well as for different sub-ranges in zenith angle are shown.
    }
    \label{fig:aeff}
\end{figure}

In this section, we provide a comparison of the IRFs of CTA and KM3NeT.
While the KM3NeT IRFs generated by ourselves are defined up to an energy of \SI{100}{PeV}, the public CTA IRFs are valid up to an energy of only $\sim$\SI{300}{TeV}.
This is because, given its limited duty cycle and need for pointing, CTA is not expected to be able to effectively measure fluxes beyond that energy.

Fig.~\ref{fig:aeff} shows a comparison of the effective areas.
The effective area of CTA rises sharply at the threshold energy of the instrument (around \SI{0.1}{TeV}), before the curve gradually flattens as \gam rays are detected more and more efficiently.
The effective area of KM3NeT for neutrinos is much lower than that of CTA for \gam rays because of the low interaction probability of neutrinos.
The increase in neutrino effective area with increasing energy reflects a corresponding increase of the interaction cross section and detection efficiency.
At the highest energies, the interaction cross section becomes large enough for the Earth to become opaque to neutrinos, leading to a decrease in the effective area for neutrinos that traverse large amounts of matter (green dashed-dotted and red dotted line in Fig.~\ref{fig:aeff}).

\begin{figure}
    \centering
    \includegraphics[width=0.99\columnwidth]{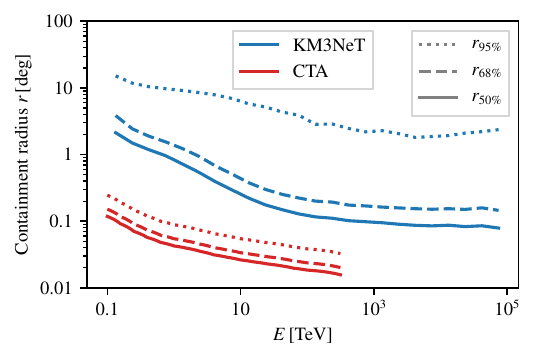}
    \caption{
      Comparison of the directional reconstruction accuracy of CTA and KM3NeT.
      Shown are the 68\% and 95\% containment radii of the PSF as a function of the true \gam-ray/neutrino energy.
      Possible differences to previous publications may arise from the finite binning of the PSF applied here.
    }
    \label{fig:psf}
\end{figure}

The angular resolutions of the two instruments -- here expressed in terms of the 50\%, 68\%, and 95\% quantiles of the respective PSFs -- are compared in Fig.~\ref{fig:psf}.
While the angular resolution of CTA is clearly superior to that of KM3NeT, the selection of track-like events for the KM3NeT analysis still leads to a median resolution of better than $0.3^\circ$ above $\sim$\SI{10}{TeV}.
The PSF strongly affects the sensitivity to point-like or marginally extended sources, as the contribution of background events increases quadratically with the radius of the source after PSF convolution.
For a comparison of the KM3NeT PSF with that of the IceCube neutrino telescope, see for example \cite{Muller2021}.

\begin{figure}
    \centering
    \includegraphics[width=0.99\columnwidth]{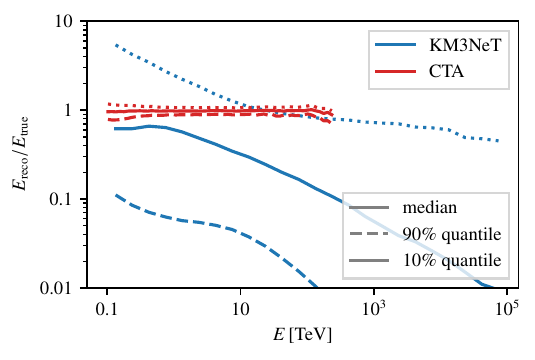}
    \caption{
      Comparison of the energy reconstruction accuracy of CTA and KM3NeT.
      Shown are the 10\%, 50\%, and 90\% quantiles of the ratio between the reconstructed and true \gam-ray/neutrino energy, as a function of the true energy.
    }
    \label{fig:edisp}
\end{figure}

Figure~\ref{fig:edisp} provides a comparison of the energy resolution of the two instruments, here indicated by the 10\%, 50\%, and 90\% quantiles of the ratio between reconstructed and true energy.
In the case of KM3NeT, muons created in muon neutrino interactions may lose energy before entering the detector or carry away energy when leaving it.
Because only the energy deposited inside the detector can reliably be estimated, the resulting reconstructed energy is on average lower than the true neutrino energy.
This effect becomes more and more prominent as the neutrino energy -- and hence the track length of the resulting muon -- increases.

\subsection{Input models}
\label{sec:models}
\begin{figure}
    \centering
    \includegraphics[width=0.99\columnwidth]{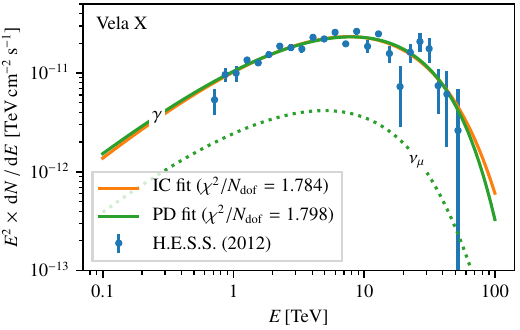}
    \caption{
        Fit of hadronic (PD) and leptonic (IC) input models for \velax.
        The muon-neutrino prediction based on the best-fit PD model is shown as dashed line.
        The data points are taken from \cite{HESS_VelaX}, and based on \SI{53}{\hour} of H.E.S.S.\ observations.
    }
    \label{fig:fit_fp_velax_main}
\end{figure}

Input models are needed for the likelihood analysis of each analysed source (cf.\ Table~\ref{tab:sources}).
During the analysis procedure a spatial model is chosen for each source according to the description found in the literature: either a uniform disk or a two-dimensional Gaussian model.
These spatial models are not varied or fitted (i.e.\ remain fixed) during the entire analysis procedure.
In addition, two different spectral models have been considered for each source in order to study the fraction of hadronic \gam-ray emission: an IC model, assuming a purely leptonic emission and a PD model, to describe a purely hadronic emission.
Both models have independently been fitted to published \gam-ray spectra of the sources (cf.\ references in Table~\ref{tab:sources}).
For the IC model, we have used the implementation of the \texttt{InverseCompton} model in the \textsc{naima} package \cite{Zabalza2015}, which provides one-zone, time-independent radiative models.
For the PD model, we have implemented a corresponding model based on the parametrisation in \cite{Kelner2006}\footnote{We are aware of more recent parametrisations such as that provided by \cite{Kafexhiu2014}, which is also used in the \texttt{PionDecay} model implementation in \textsc{naima}. That parametrisation, however, does not provide a prediction for the expected neutrino flux, required for our analysis. The focus of this study lying on the technical feasibility of a joint \gam-ray/neutrino analysis, the choice of parametrisation for the hadronic model is not relevant here.}.
For both the IC and PD models, we assume a power-law model with an exponential cut-off for the primary electron and proton distributions,
\begin{linenomath*}
\begin{equation}
    \label{eq:ecpl}
    \Phi(E) = A \cdot \left(\frac{E}{E_0}\right)^{-\Gamma} \exp\left[-\left(\frac{E}{E_\mathrm{cut}}\right)^\beta\right]\,,
\end{equation}
\end{linenomath*}
where $A$ denotes the amplitude, $E_0$ the reference energy, $\Gamma$ the spectral index, $E_{\mathrm{cut}}$ the cut-off energy, and $\beta$ the cut-off strength.
For each source and for both the IC and PD models, we have adjusted the amplitude, spectral index, and cut-off energy using a simple $\chi^2$ fit, keeping the reference energy and cut-off strength fixed (at $E_0=\SI{10}{TeV}$ and $\beta=1$, respectively).
The fit for \velax is shown in Fig.~\ref{fig:fit_fp_velax_main}, whereas those for the other sources can be found in \ref{sec:appendix_model_fits}.
Both models describe the \gam-ray flux equally well, illustrating the difficulty to distinguish between the leptonic and hadronic scenarios based on \gam-ray data alone.
We note that the addition of lower-energy radio or X-ray data would further constrain the fit, but regard this as beyond the scope of this work.

\subsection{Generation of pseudo data sets}
\label{sec:dataset_gen}
\begin{figure}
    \centering
    \includegraphics[width=0.99\columnwidth]{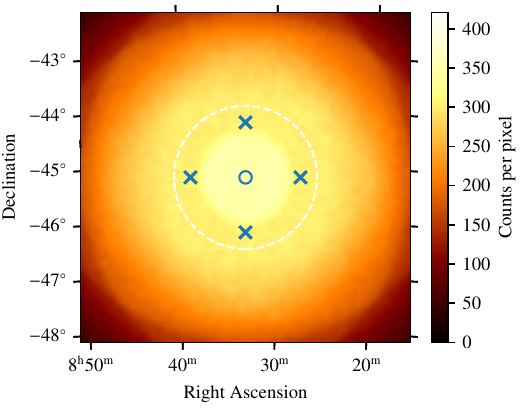}
    \caption{
        Counts map of a pseudo CTA data set based on the PD model for \velax with 200~hours of observation time.
        The counts are Poisson-randomised based on the model prediction for \velax (modelled as a disk with radius $0.8^\circ$) and the residual hadronic background, summed over all energies and smoothed with a 0.05$^\circ$ Gaussian.
        The blue circle and `$\times$' markers denote the source and pointing positions, respectively.
        The white dashed circle shows the source region for which the counts spectra shown in Fig.~\ref{fig:counts_spec_cta_velax} have been extracted.
    }
    \label{fig:counts_map_cta_velax}
\end{figure}

\begin{figure}
    \centering
    \includegraphics[width=0.99\columnwidth]{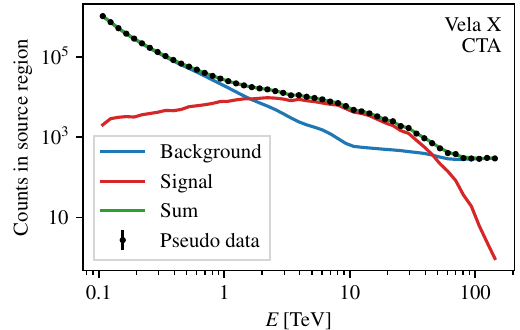}
    \caption{
        Counts spectra for the \velax CTA PD data set, extracted for a region encompassing the source (cf.\ Fig.~\ref{fig:counts_map_cta_velax}).
        The coloured lines denote the number of predicted counts within the source region for an observation time of 200 hours.
        The black data points visualise one random Poisson realisation, drawn from the model predictions.
    }
    \label{fig:counts_spec_cta_velax}
\end{figure}

With the IRFs and input models in hand, we generated 100 pseudo data sets for each source and each instrument, both for the PD and IC models.
We note that for both instruments, we do not take into account diffuse astrophysical \gam-ray or neutrino emission that is unrelated to the source itself.
For the CTA data sets we used an analysis setup with 16 energy bins per decade between \SI{0.1}{TeV} and \SI{154}{TeV} and spatial bins of $0.02^\circ\times 0.02^\circ$ size.
For each pseudo data set, we assumed a total observation time of 200~hours, split equally between four pointing positions with $1^\circ$ offset with respect to the source position. 
The predicted number of source and background events are summed for each pixel and Poisson-distributed random counts are drawn based on those values.
As an example, in Figs.~\ref{fig:counts_map_cta_velax} and~\ref{fig:counts_spec_cta_velax}, we show projections of one generated pseudo CTA data set based on the PD model for the source \velax onto the spatial and energy axes, respectively.

\begin{figure}
    \centering
    \includegraphics[width=0.99\columnwidth]{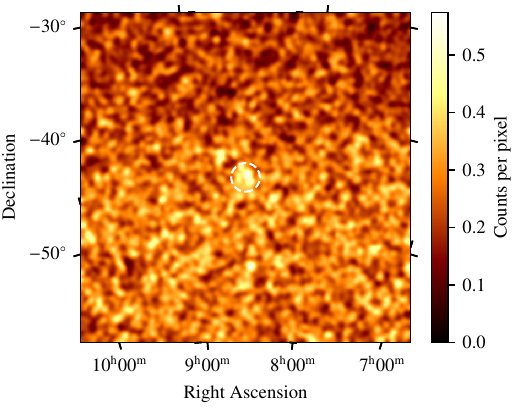}
    \caption{
        Counts map of a pseudo KM3NeT data set based on the PD model for \velax with 10~years of observation time.
        The counts are Poisson-randomised based on the model prediction for \velax, summed over all energies and smoothed with a 0.25$^\circ$ Gaussian.
        The white dashed circle shows the source region for which the counts spectra shown in Fig.~\ref{fig:counts_spec_km3net_velax} have been extracted (same region as in Fig.~\ref{fig:counts_map_cta_velax}).
    }
    \label{fig:counts_map_km3net_velax}
\end{figure}

\begin{figure}
    \centering
    \includegraphics[width=0.99\columnwidth]{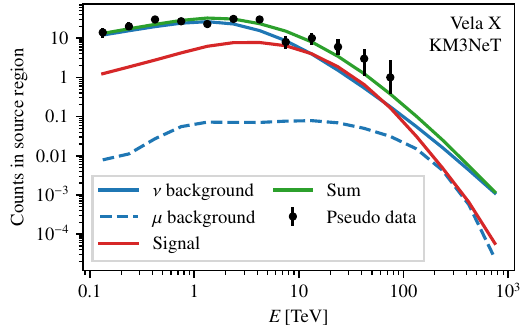}
    \caption{
        Counts spectra for the \velax KM3NeT PD data set, extracted for a region encompassing the source (cf.\ Fig.~\ref{fig:counts_map_km3net_velax}).
        The coloured lines denote the number of predicted counts within the source region for an observation time of 10~years.
        The black data points visualise one random Poisson realisation, drawn from the model predictions.
    }
    \label{fig:counts_spec_km3net_velax}
\end{figure}

A similar procedure has been used to generate the KM3NeT pseudo data sets, for which we assume a total detector operation time of 10~years.
For each zenith angle bin (cf.\ Fig.~\ref{fig:visibility}), we generate an observation set evaluating the associated IRFs, according to the corresponding fraction of the total observation time.
This results in 6 or 7 observation sets for each source, depending on the visibility.
The exposure and expected background for every data set are computed in equatorial coordinates by integrating over time the respective IRFs (defined in terms of the local zenith angle).
Data are binned using spatial pixels of $0.1^\circ\times 0.1^\circ$ size and 4 bins per decade in energy, between \SI{100}{GeV} and \SI{1}{PeV}.
The IRFs are evaluated using a finer binning in energy, with 16 bins per decade between \SI{100}{GeV} and \SI{10}{PeV}.
Several tests have been performed with finer zenith, energy, and spatial binning, yielding consistent results and no significant change in sensitivity.
An example of a KM3NeT pseudo data set is visualised in Figs.~\ref{fig:counts_map_km3net_velax} and~\ref{fig:counts_spec_km3net_velax}, respectively.

\subsection{Likelihood analysis}
\label{sec:analysis}
The analysis is performed using the binned likelihood formalism implemented in the \gp package\footnote{We note that analyses carried out with the analysis tools normally employed by the KM3NeT Collaboration are typically performed using an unbinned likelihood formalism, which can enhance the sensitivity. An unbinned analysis is however not yet implemented in \gp.}.
Leptonic and hadronic models are fitted to the generated pseudo data sets\footnote{The IC model, predicting \gam rays but no neutrinos, is of course fitted to the CTA data sets only.} by minimising the `Cash statistic' \cite{Cash1979}
\begin{linenomath*}
\begin{equation}
\label{eq:cash}
    \mathcal{C}(\xi) = -2\ln\mathcal{L}(\xi)\,,
\end{equation}
\end{linenomath*}
where
\begin{linenomath*}
\begin{equation}
    \mathcal{L}(\xi) = \prod_{i=1}^N \mathcal{P}(n_i | \nu_i(\xi))
\end{equation}
\end{linenomath*}
denotes the total likelihood to observe the generated data, and
\begin{linenomath*}
\begin{equation}
    \mathcal{P}(n_i | \nu_i(\xi)) = \frac{\nu_i^{n_i}(\xi)}{n_i!} \cdot \exp(-\nu_i(\xi))
\end{equation}
\end{linenomath*}
is the Poisson probability to measure $n_i$ events in pixel $i$, given a model prediction $\nu_i(\xi)$ that depends on the model parameters $\xi$ and is computed taking into account the IRFs of the instruments.
Several data sets can be simultaneously fitted by multiplying their respective likelihood values.
Because the data sets are analysed with the same IRFs that were also used to create them, systematic uncertainties related to the generation of the IRFs are not taken into account here.

Confidence intervals for specific model parameters can be obtained by means of a profile likelihood scan.
In the scan, the parameter of interest $x$ is consecutively fixed to values $x_\mathrm{scan}$ around the optimum value $\hat{x}$, while the other model parameters are optimised in each step.
A confidence interval can then be derived from the difference in `test statistic',
\begin{linenomath*}
\begin{equation}\label{eq:ts}
    \Delta\mathrm{TS}(x_\mathrm{scan}) = -2 \ln\left(\frac{\mathcal{L}(x_\mathrm{scan},\xi)}{\mathcal{L}(\hat{\xi})}\right)\,,
\end{equation}
\end{linenomath*}
where $\hat{\xi}$ are the parameter values for which $\mathcal{L}$ is maximal.
As a result of the re-optimisation of the other parameters $\xi$ the test statistic is effectively only a function of the parameter of interest.

\subsection{Constraining the hadronic contribution}
\label{sec:computation-of-cl}
In this work, we derive credible intervals for the contribution of hadronic emission processes (i.e.\ the PD model) to the total \gam-ray emission of the investigated sources.
To this end, we simultaneously fit a PD model and an IC model to each of the pseudo data sets (i.e.\ the total \gam-ray emission is given by the sum of the two models), which were generated with either the PD model or the IC model as input.
The free parameters $\xi$ of this composite model are the parameters $\xi_\mathrm{p}$ describing the proton population and the parameters $\xi_\mathrm{e}$ for the electron distribution (cf.\ Eq.~\ref{eq:ecpl}).
Our parameter of interest is then the `hadronic fraction'
\begin{linenomath*}
\begin{equation}
    f(\xi) = \frac{I_\mathrm{had}(\xi_\mathrm{p})}{I_\mathrm{had}(\xi_\mathrm{p}) + I_\mathrm{lep}(\xi_\mathrm{e})}\,,
\end{equation}
\end{linenomath*}
where $I_\mathrm{had}$ and $I_\mathrm{lep}$ denote the integrated \gam-ray flux between \SI{100}{GeV} and \SI{100}{TeV} for the best-fit hadronic (PD) and leptonic (IC) model, respectively.
Since the hadronic fraction is not a direct parameter of the model and we cannot simply fix it to certain values, we add a penalty term $P$ to the Cash statistic,
\begin{linenomath*}
\begin{equation}
\label{eq:cash_tot}
    \mathcal{C}_\mathrm{tot} = \mathcal{C}(\xi) + P(\xi,f_\mathrm{scan})\,,
\end{equation}
\end{linenomath*}
where
\begin{linenomath*}
\begin{equation}
    P(\xi,f_\mathrm{scan}) = A_p\cdot(f(\xi)-f_\mathrm{scan})^2 / \Delta f^2
\end{equation}
\end{linenomath*}
and $f_\mathrm{scan}$ is the value of $f$ that we want to probe.
This penalty term allows us to fully re-optimise the model (i.e.\ all its direct parameters $\xi$) while maintaining a hadronic fraction close to $f_\mathrm{scan}\pm\Delta f$.
It is thus mostly a technical tool that enables us to carry out profile likelihood scans for $f$.
We scan 21~values equally spaced between~0 and~1.\footnote{
  We note that the contribution of the penalty term is not included in the further evaluation of the likelihood.
  Furthermore, in order to take into account the small possible variations of $f$ around $f_\mathrm{scan}$, the exact hadronic fractions resulting from the optimised model are used in the subsequent analysis.
}
Values of $A_p=0.1$ and $\Delta f=0.01$ were empirically found to lead to a strong-enough constraint -- that is, to ensure that the allowed variation in $f$ is small compared to the spacing of the scan values -- while yielding stable results.
In the following, we denote with $\hat{\xi}_\mathrm{scan}$ the best-fit parameter values for a given $f_\mathrm{scan}$, and with $\hat{\xi}$ the parameter values corresponding to the overall best fit (i.e.\ without a penalty term that constrains $f$).

In the limit of sufficient statistics and for parameter values far enough from parameter boundaries, Wilk's theorem \cite{Wilks1938} states that $\Delta\mathrm{TS}$ follows a $\chi^2$ distribution and can therefore directly be used to deduce confidence intervals \cite{Workman2022}.
However, as the parameter $f$ is bounded between 0~and~1, we cannot invoke Wilk's theorem here.
An alternative method would be to derive the expected distribution of $\Delta\mathrm{TS}$ by generating and fitting a large number ($\gg$100) of pseudo data sets.
Unfortunately, the profile likelihood scan with full re-optimisation of all model parameters is rather computing-intensive, implying that this approach is also not feasible here.
We therefore adopt a Bayesian approach, in which we infer a posterior probability density function (PDF) $\Phi(f)$ from the likelihood ratio $\mathcal{L}(\hat{\xi}_\mathrm{scan})/\mathcal{L}(\hat{\xi})$.
By definition (cf.\ Eq.~\ref{eq:ts}), 
\begin{linenomath*}
\begin{equation}
    \frac{\mathcal{L}(\hat{\xi}_\mathrm{scan})}{\mathcal{L}(\hat{\xi})} = \frac{\mathcal{L}(f_\mathrm{scan},\xi)}{\mathcal{L}(\hat{\xi})} = \exp\left(-\frac{1}{2}\Delta \mathrm{TS}(f_\mathrm{scan})\right)\,,
\end{equation}
\end{linenomath*}
which we use to derive the PDF as
\begin{linenomath*}
\begin{equation}
\label{eq:pdf}
    \Phi(f) = c\cdot \exp\left(-\frac{1}{2}\Delta\mathrm{TS}(f)\right)\,,
\end{equation}
\end{linenomath*}
where $f\equiv f(\hat{\xi}_\mathrm{scan})$ and $c$ is a normalisation constant that can be determined by requiring $\int_0^1 \Phi(f)\,\mathrm{d}f=1$.
To obtain a smooth curve, we fit the $\Delta\mathrm{TS}$ values obtained from the profile likelihood scan with a cubic spline function.
A central Bayesian credible interval for $f$ can then be derived by integrating the PDF around the best-fit value up to a certain probability (e.g.\ 68\% or 90\%) -- this is also known as the construction of a highest posterior density interval \cite{Berger1980}.
We assume a flat prior distribution for $f$ in this procedure.
An exemplary PDF together with its 90\% credible interval is shown in Fig. \ref{fig:pdf} in Appendix~\ref{sec:appendix_pdf}.

\section{Results}
\label{sec:results}
In this section, we will introduce the different analysis scenarios we have considered (Sect.~\ref{sec:ana_scenarios}), present the results (Sect.~\ref{sec:ana_results}) and discuss their implications (Sect.~\ref{sec:discussion}).

\subsection{Analysis scenarios}
\label{sec:ana_scenarios}
For every source, input model (leptonic or hadronic), and pseudo data set, we begin by fitting the PD and IC model to only the CTA data sets.
Here we obtain the optimal parameters $\xi_\mathrm{p}$ and $\xi_\mathrm{e}$ for each model, which serve as starting parameters for the following, more complicated, composite model.
In total we perform the analysis in three different scenarios, each with the goal of recovering the hadronic fraction and its uncertainty (cf.\ Sect.~\ref{sec:computation-of-cl}).
The first scenario is a ``CTA only'' analysis, in which we analyse only the \gam-ray data provided by CTA and ignore the KM3NeT neutrino data.
This scenario is intended to illustrate to which degree CTA alone can differentiate between the two models, based on the \gam-ray energy spectrum.
As mentioned before, we perform a profile likelihood scan of the hadronic fraction by re-optimising the composite model (the sum of PD and IC model) at different ratios of hadronic to leptonic \gam-ray flux prediction.
Second, we perform a ``KM3NeT only'' analysis, where we include only the neutrino data provided by KM3NeT.
However, as the primary CR energy spectrum can presently not be measured well with neutrinos alone, we constrain the parameters of the PD model to be consistent with those derived in the fit of the pure PD model to the CTA data sets, using the same concept of penalty terms as for the hadronic fraction $f$ (cf.\ the previous section).
Specifically, for each free parameter $p$ of the PD model, we add a term $(p-\hat{p})^2/\Delta\hat{p}^2$, where $\hat{p}$ and $\Delta\hat{p}$ are the best-fit parameter value and its uncertainty, respectively.
Thus, the second scenario shows how well the leptonic and hadronic scenario can be distinguished with KM3NeT data if the primary CR energy spectrum is known to a certain degree.
As the IC model is irrelevant for the neutrino flux, the hadronic fraction now corresponds to the ratio of \gam-ray flux expected from the fitted proton distribution to the total \gam-ray flux measured with CTA.
Finally, in a third scenario we combine the \gam-ray and neutrino data and perform a joint analysis of the CTA and KM3NeT data sets.
As the primary CR energy spectra are now directly constrained by the CTA data, the prior terms added for the second scenario are removed again.
This scenario demonstrates the benefits of the combined analysis.
We note that, if the best-fit models obtained in the three scenarios are similar, a combination of the profile likelihood scans performed in the first two scenarios will lead to the same constraints on the hadronic fraction as the combined analysis in the third scenario.
While this is often the case for the relatively simple models employed here, the situation can be different for more complex models, for which the combined analysis may be able to break degeneracies between model parameters.

\subsection{Analysis results}
\label{sec:ana_results}
\begin{figure}
    \centering
    \includegraphics[width=0.99\columnwidth]{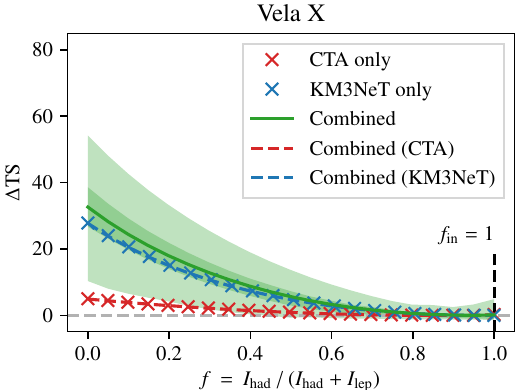}
    \caption{
        Profile likelihood scan results for \velax and a hadronic input model ($f_\mathrm{in}=1$).
        The displayed curves show the average of the curves obtained for the 100 generated pseudo data sets.
        The red and blue `$\times$' markers correspond to the analysis scenarios ``CTA only'' and ``KM3NeT only'', respectively.
        The green line shows the result for the combined analysis, together with 68\% and 95\% quantile intervals to indicate the statistical spread.
        The dashed red and blue lines show the respective contributions of the CTA and KM3NeT data sets to the combined result.
    }
    \label{fig:scan_velax}
\end{figure}

As an example, we show in Fig.~\ref{fig:scan_velax} average profile likelihood scans of the hadronic fraction $f$ for \velax, where the hadronic PD model has been used as input when generating the pseudo data sets.
Corresponding plots for the other sources and for a purely leptonic (IC) input model can be found in \ref{sec:appendix_other_scans}.
By definition, $\Delta\mathrm{TS}=0$ at the minimum of the curves, which (as expected) occurs at the value of $f$ that corresponds to the input model (i.e.\ $f_\mathrm{in}=0$ for a purely leptonic input model and $f_\mathrm{in}=1$ for a purely hadronic input model).
Moving away from the minimum, larger values of $\Delta\mathrm{TS}$ imply a stronger rejection of the corresponding hadronic fraction $f$.

\begin{figure}
    \centering
    \includegraphics[width=0.99\columnwidth]{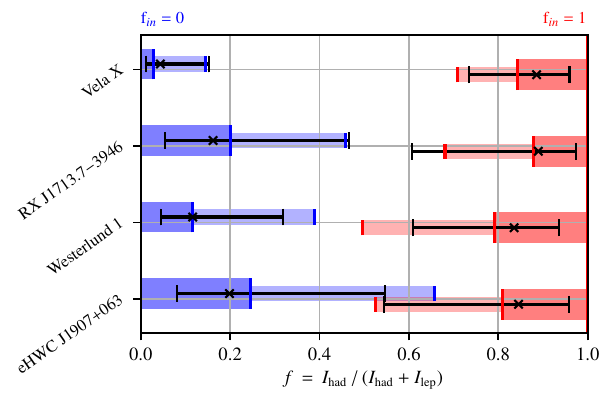}
    \caption{
        Summary of results for all sources and input models.
        The blue and red horizontal bars show the 68\% and 90\% quantile intervals of the distribution of best-fit values for the leptonic ($f_\mathrm{in}=0$) and hadronic ($f_\mathrm{in}=1$) input models, respectively.
        The black bars indicate the average position and sizes of the 68\% credible intervals derived for each input model.
    }
    \label{fig:limits-all}
\end{figure}

In Fig.~\ref{fig:limits-all}, we display the 68\% and 90\% quantile intervals of the distribution of best-fit values of the hadronic fraction $f$ for all 100 pseudo data sets.
Furthermore, as described in Sect.~\ref{sec:computation-of-cl}, we obtain credible intervals for $f$ from the profile likelihood scans.
The average expected 68\% credible intervals are indicated by the black bars.
We note that, for individual pseudo data sets, it is possible that the 68\% credible interval does not contain the true input value of $f$.
Hence, this is also the case for the averaged interval: this is an artefact caused by the fact that the input values lie at the boundaries of the allowed values for $f$.

\subsection{Discussion}
\label{sec:discussion}
It is evident that in essentially all cases, the ``CTA only'' and ``KM3NeT only'' scenarios yield results that correspond exactly to the contributions of the two instruments in the combined analysis.
That is, in Figs.~\ref{fig:scan_velax}, \ref{fig:scans1}, and \ref{fig:scans2}, the red and blue crosses fall on top of the red and blue dashed lines, respectively.
This means that consistent source models are fitted in all three scenarios and implies that, in principle, a sensitivity identical to that of the combined analysis can be achieved by combining the results of the profile likelihood scans of the single-instrument analyses.
We note, however, that this only holds because we use the same source models in all analysis scenarios, and because we incorporate the CTA constraints on the primary CR spectrum in the ``KM3NeT only'' analysis.
In practice, ensuring this consistency between two separate analyses may not always be possible -- in particular when more sophisticated source models are considered, which could, for example, feature different spatial models for the leptonic and hadronic cases.
The combined analysis, on the other hand, naturally ensures that the \gam-ray and neutrino data are described with the same physical models.
Furthermore, a combined analysis enables a consistent incorporation of systematic uncertainties: for example, it would be possible to add to the analysis a nuisance parameter that modifies the relative energy scale between the two instruments -- something that would be much more difficult to take into account when combining the results of two single-instrument analyses.

Investigating the profile likelihood scans for the different sources, we note that the curves obtained from the KM3NeT data sets are usually very similar for the leptonic and hadronic input model, except being flipped horizontally, and often yield stronger constraints on the hadronic fraction $f$ compared to the CTA curves.
This is not unexpected, as the predicted neutrino fluxes differ fundamentally between the leptonic model (no neutrinos) and the hadronic model (neutrino flux approximately equal to \gam-ray flux), whereas the predicted \gam-ray fluxes can look very similar (cf.\ Figs.~\ref{fig:fit_fp_velax_main},~\ref{fig:fit_fp_other}).
This is especially true here, as we use the same spatial models for the leptonic and hadronic scenario, implying that for CTA any separation power between the models originates from differences in the predicted \gam-ray spectra.
For more realistic models that also differ in their morphology, the CTA data will lead to better constraints than achieved here -- the presented curves should therefore not be regarded as a generally valid estimate of the CTA sensitivity to differentiate between leptonic and hadronic models.
The KM3NeT results, on the other hand, are more representative of the true expected sensitivity of the instrument.
Here, the achieved constraints on $f$ mostly depend on the hardness of the fitted input spectra, that is, the predicted \gam-ray flux at the highest energies ($>\SI{10}{TeV}$).
The large statistical spread in the achieved constraints (indicated by the green bands in Figs.~\ref{fig:scan_velax}, \ref{fig:scans1}, and \ref{fig:scans2} and by the blue and red bars in Fig.~\ref{fig:limits-all}) is caused by the small number of neutrinos that is expected to be detectable with KM3NeT, even for the most promising sources.

In the following, we briefly discuss the results obtained for the individual studied sources.
A more general assessment of the sensitivity of KM3NeT to the sources studied in this work, except for \wld, can be found in \cite{KM3NeT_PSsens_2019}.

\emph{\velax} --- For \velax we found the strongest discrimination potential for all sources investigated in this study.
This is largely due to its full visibility for KM3NeT and its large \gam-ray flux in the \SIrange{10}{100}{TeV} range.
The large $\Delta\mathrm{TS}$ values obtained with the CTA data sets for the leptonic input model are a result of the strongly curved measured \gam-ray spectrum, which is easier to reproduce with an IC model.
While it appears clear that a purely hadronic origin of the \gam-ray emission is already excluded by studies in the \gam-ray domain (e.g.\ \cite{HESS_VelaX,Tibaldo2018}), our results indicate that, in the case of a purely leptonic origin, a combined analysis of \gam-ray and neutrino data may help in ruling out a potential small contribution to the \gam-ray flux from hadronic processes.
Specifically, the obtained average 68\% credible interval constrains the hadronic fraction $f$ to below 15\%.

\emph{\rxj} --- Compared to \velax, the spectrum of \rxj exhibits less curvature, which reflects in the flat $\Delta\mathrm{TS}$ curves obtained with CTA for this source.
As the expected neutrino flux is slightly lower than for \velax, the resulting constraints on the hadronic fraction $f$ are weaker and the statistical spread is larger.
Nevertheless, as the physical processes responsible for the \gam-ray emission from \rxj are not yet totally clear, the addition of neutrino data from KM3NeT may yield important new constraints.

\emph{\wld} --- Compared to the other sources studied in this work, the stellar cluster \wld has a relatively hard energy spectrum without a strong cut-off, which leads -- in a hadronic scenario -- to an expected flux of neutrinos that is detectable with KM3NeT.
Therefore, despite \wld being a significantly extended source (which implies a larger background contamination), a combined analysis can, on average, distinguish between the leptonic and hadronic models investigated here.
However, as the statistical spread is very large, the limits obtained from the different pseudo data sets differ strongly.

\emph{\ehwc} --- In contrast to the other sources, \ehwc is modelled using a Gaussian spatial model and has the largest 68\% flux containment radius.
Because it is furthermore visible for only 54\% of the time for KM3NeT, the constraints obtained for this source are the weakest in this study.
This is reflected in the overlapping 90\% quantile intervals of the fitted hadronic fractions $f$ for the leptonic and hadronic input models (see Fig.~\ref{fig:limits-all}).
For the case of a hadronic input model, the CTA data sets yield some separation power as the IC model, being intrinsically curved, cannot reproduce well the measured spectrum which shows only a small curvature.
Nevertheless, we conclude that even with \SI{200}{\hour} of CTA data and \SI{10}{yr} of KM3NeT data, a differentiation between a leptonic and a hadronic scenario is challenging.

\section{Conclusion}
\label{sec:conclusion}
In this work, we explore the prospects of a combined analysis of \gam-ray and neutrino measurements with the upcoming CTA and KM3NeT facilities.
To this end, we demonstrate the capability of the open-source \gam-ray analysis package \gp to also analyse data from neutrino telescopes such as KM3NeT, even though these observe the sky in a conceptually different way compared to \gam-ray telescope arrays such as CTA.\footnote{
  We note that, since the initiation of this work, several improvements to both the \gp package and the `GADF' data format specifications have further eased the analysis of data from wide-field instruments, which besides neutrino telescopes also include \gam-ray observatories that employ particle sampling detectors, like the HAWC experiment \cite{HAWC2017,HAWC2022}.
}
For the combined analysis, we use publicly available IRFs for CTA and generate custom IRFs for KM3NeT.
We fit measured \gam-ray spectra of four prototypical Galactic \gam-ray sources to obtain both a leptonic model and a hadronic model.
Using either of these models as well as the IRFs as input, we calculate the expected \gam-ray and neutrino flux and generate pseudo data sets for both CTA and KM3NeT for all sources.
Using a binned likelihood analysis approach, we derive from these pseudo data sets average expected constraints on the contribution of hadronic emission processes to the total emission.

We find that our combined analysis approach enables a consistent modelling of the \gam-ray and neutrino observations.
Due to simplifying assumptions (e.g.\ fixed spatial models), the sensitivity to constrain the physical mechanism responsible for the \gam-ray emission is often dominated by the KM3NeT data; we note that this will not generally be the case when employing more sophisticated models.
For sources that emit a sufficiently large flux of \gam rays at high energies ($>\SI{10}{TeV}$), our results indicate that a combined analysis of \SI{200}{\hour} of CTA and \SI{10}{yr} of KM3NeT data will be able to differentiate between a leptonic and a hadronic emission scenario.

This work constitutes the first demonstration of a general framework for jointly analysing event-level data from \gam-ray and neutrino telescopes in a likelihood formalism.
Here, we have applied it to simulated data from the CTA and KM3NeT observatories, for a selection of Galactic \gam-ray sources, with the aim of deriving expected constraints on the contribution of hadronic emission processes.
This approach may be complemented with others, for example the analysis of multi-wavelength data (e.g.\ in the radio or X-ray domain).
On the other hand, the combination of \gam-ray and neutrino data can be applied to many more questions.
For example, there is increasing evidence that active galactic nuclei -- a prominent extragalactic \gam-ray source class -- are also neutrino sources, see for example \cite{IceCube2018,IceCube2022}.
In this regard, it is worth noting that the data recorded with both the CTA and KM3NeT observatories will be made publicly available -- opening a bright future for multi-messenger analyses.

We release along with this paper a software repository that provides the necessary code, in the form of \textsc{Jupyter} notebooks, to reproduce the presented results and figures~\cite{Unbehaun2023}.
It can be found at the following URL:\\\href{https://zenodo.org/record/8298464}{https://zenodo.org/record/8298464}.

\begin{acknowledgements}
The KM3NeT Collaboration acknowledges the financial support of the funding agencies:
Agence Nationale de la Recherche (contract ANR-15-CE31-0020), Centre National de la Recherche Scientifique (CNRS), Commission Europ\'eenne (FEDER fund and Marie Curie Program), LabEx UnivEarthS (ANR-10-LABX-0023 and ANR-18-IDEX-0001), Paris \^Ile-de-France Region, France;
Shota Rustaveli National Science Foundation of Georgia (SRNSFG, FR-22-13708), Georgia;
The General Secretariat of Research and Innovation (GSRI), Greece;
Istituto Nazionale di Fisica Nucleare (INFN), Ministero dell'Universit\`a e della Ricerca (MIUR), PRIN 2017 program (Grant NAT-NET 2017W4HA7S) Italy;
Ministry of Higher Education, Scientific Research and Innovation, Morocco, and the Arab Fund for Economic and Social Development, Kuwait;
Nederlandse organisatie voor Wetenschappelijk Onderzoek (NWO), the Netherlands;
The National Science Centre, Poland (2021/41/N/ST2/01177);
National Authority for Scientific Research (ANCS), Romania;
Grants PID2021-124591NB-C41, -C42, -C43 funded by MCIN/AEI/ 10.13039/501100011033 and, as appropriate, by ``ERDF A way of making Europe'', by the ``European Union'' or by the ``European Union NextGenerationEU/PRTR'', Programa de Planes Complementarios I+D+I (refs.\ ASFAE/2022/023, ASFAE/2022/014), Programa Prometeo (PROMETEO/2020/019) and GenT (refs.\ CIDEGENT/2018/034, /2019/043, /2020/049, /2021/23) of the Generalitat Valenciana, Junta de Andaluc\'{i}a (ref. SOMM17/6104/UGR, P18-FR-5057), EU: MSC program (ref. 101025085), Programa Mar\'{i}a Zambrano (Spanish Ministry of Universities, funded by the European Union, NextGenerationEU), Spain;
The European Union's Horizon 2020 Research and Innovation Programme (ChETEC-INFRA - Project no. 101008324).

T.~Unbehaun, L.~Mohrmann, and S.~Funk are members of the CTA Consortium.
The paper has been reviewed by the CTA Consortium Speakers and Publications Office.

This research has made use of the ``Prod 5'' CTA instrument response functions provided by the CTA Consortium and Observatory, see \url{https://www.cta-observatory.org/science/cta-performance} for more details.
This research made use of the \textsc{Astropy} (\url{http://www.astropy.org}; \cite{Robitaille2013,PriceWhelan2018}) and \textsc{matplotlib} (\url{http://www.matplotlib.org}; \cite{Hunter2007}) software packages.
The authors gratefully acknowledge the compute resources and support provided by the Erlangen National High Performance Computing Center (NHR@FAU).
The publication of this analysis in the open science regime under FAIR principles \cite{ChueHong2022} is pursued in the EOSC Future (\url{https://eoscfuture.eu}) and ESCAPE (\url{https://projectescape.eu}) projects that received funding from the European Union’s Horizon Europe and 2020 programme under Grant Agreements no.\ 101017536 and 824064, respectively.
\end{acknowledgements}

\begin{appendix}

\section{Additional input model fits}
\begin{figure*}[hb]
  \centering
  \subfigure[]{
    \includegraphics[width=0.48\textwidth]{img/fit_fp_vela_x}
    \label{fig:fit_fp_vela_x}
  }
  \subfigure[]{
    \includegraphics[width=0.48\textwidth]{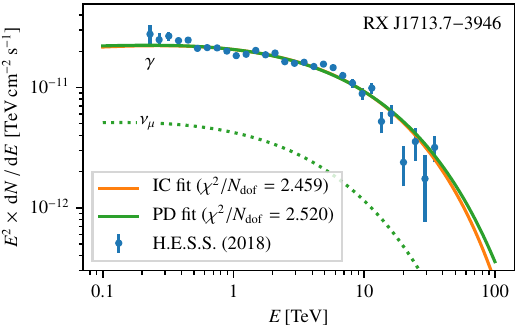}
    \label{fig:fit_fp_rxj}
  }\\
  \subfigure[]{
    \includegraphics[width=0.48\textwidth]{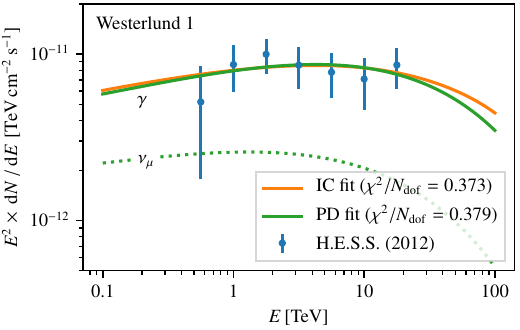}
    \label{fig:fit_fp_wd1}
  }
  \subfigure[]{
    \includegraphics[width=0.48\textwidth]{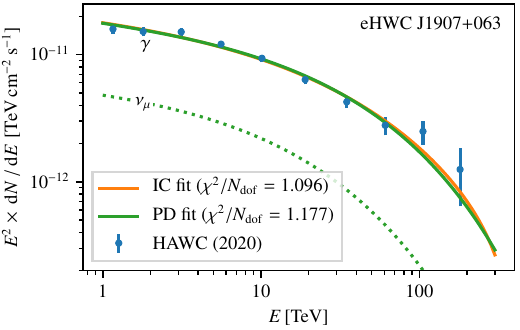}
    \label{fig:fit_fp_j1908}
  }
  \caption{
    Summary of the \gam-ray input models used for this study.
    (a) \velax (already shown in Fig.~\ref{fig:fit_fp_velax_main}), flux points taken from \cite{HESS_VelaX}; (b) \rxj, flux points taken from \cite{HESS_RXJ}; (c) \wld, flux points taken from \cite{HESS_Wd1}; (d) \ehwc, flux points taken from \cite{HAWC2020}.
    For every source, both a PD and an IC model are fitted to the flux points.
    The goodness-of-fit of the $\chi^2$ fits is indicated in the legend. The prediction of the muon-neutrino flux based on the best-fit PD model is shown as a dotted line.
  }
  \label{fig:fit_fp_other}
\end{figure*}

Input model fits for all sources are shown in Fig.~\ref{fig:fit_fp_other}.
The best-fit parameter values are summarised in Table~\ref{tab:input_model_pars}.
In all fits, we have adopted a distance to the source as listed in Table~\ref{tab:sources}.
For the PD model, we have assumed an ambient gas density of \SI{1}{\per\cubic\centi\meter}.
While this density is certainly not a valid assumption for all of the studied sources, this has no impact on the analysis results presented in Sect.~\ref{sec:ana_results}, as the sum of the \gam-ray emission predicted by the leptonic and hadronic models is always constrained to match the total observed flux.
Different gas densities would thus only imply different normalisations of the primary particle spectra, but have no effect on the derived constraints on the hadronic fraction.

\label{sec:appendix_model_fits}

\begin{table*}[hb]
  \centering
  \caption{Best-fit parameter values of input models.}
  \label{tab:input_model_pars}
  \begin{tabular}{ccccccc}
    \hline\hline
    Model & Parameter & Unit & \velax$^\mathrm{a}$ & \rxj & \wld & \ehwc\\
    \hline
    \multirow{3}{*}{PD}
        & $A$ & \SI{e35}{\per\electronvolt} &
            \num{4(45)e-5} & $1.50\pm 0.07$ & $6.2\pm 1.5$ & $6.4\pm 0.5$ \\
        & $\Gamma$ & -- & 
            $-4.9\pm 9.9$ & $2.00\pm0.08$ & $1.92\pm 0.35$ & $2.23\pm 0.09$ \\
        & $E_\mathrm{cut}$ & \si{TeV} &
            $14\pm 22$ & $110\pm30$ & $890\pm 2340$ & $540\pm 220$ \\
    \hline
    \multirow{3}{*}{IC}
        & $A$ & \SI{e33}{\per\electronvolt} &
            $0.0059\pm 0.0008$ & $0.283\pm 0.013$ & $1.17\pm 0.19$ & $1.20\pm 0.10$ \\
        & $\Gamma$ & -- & 
            $0.58\pm 0.40$ & $2.74\pm 0.12$ & $2.65\pm 0.49$ & $3.15\pm 0.12$ \\
        & $E_\mathrm{cut}$ & \si{TeV} &
            $24\pm 6$ & $49\pm 12$ & $960\pm 11200$ & $380\pm 290$ \\
    \hline
  \end{tabular}

  {\raggedright
    See Eq.~\ref{eq:ecpl} for a definition of the parameter values.\\
    $^\mathrm{a}$ The value of $\Gamma$ for the PD model for \velax is extreme. Furthermore, the large uncertainties of all parameters of this model indicate a strong correlation between them. These results reflect that the \gam-ray emission of \velax is likely dominantly of leptonic origin, and the strongly curved spectrum can be fitted with the PD model with difficulty only.\par
  }
\end{table*}

\section{Likelihood scan results for all sources}
\label{sec:appendix_other_scans}
\begin{figure*}[hb]
  \centering
  \subfigure{
    \includegraphics[width=0.48\textwidth]{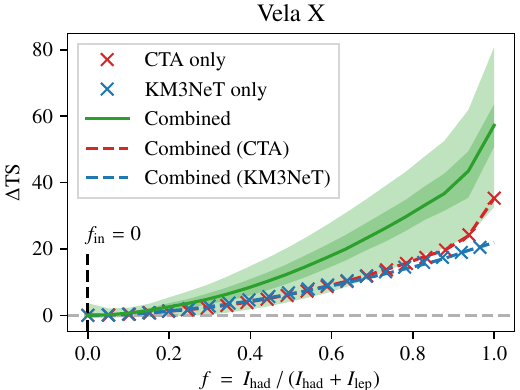}
  }
  \subfigure{
    \includegraphics[width=0.48\textwidth]{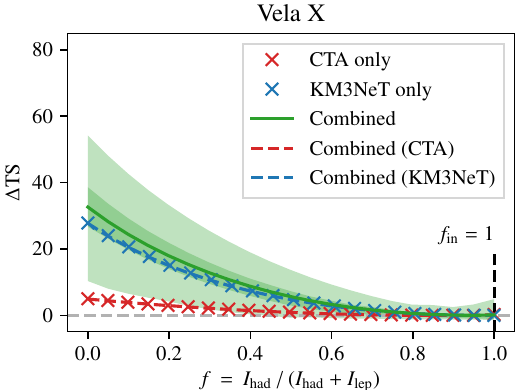}
  }\\
  \subfigure{
    \includegraphics[width=0.48\textwidth]{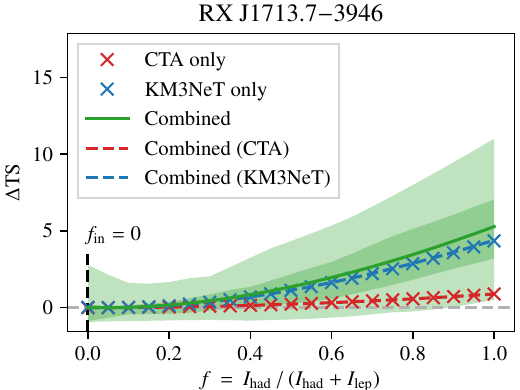}
  }
  \subfigure{
    \includegraphics[width=0.48\textwidth]{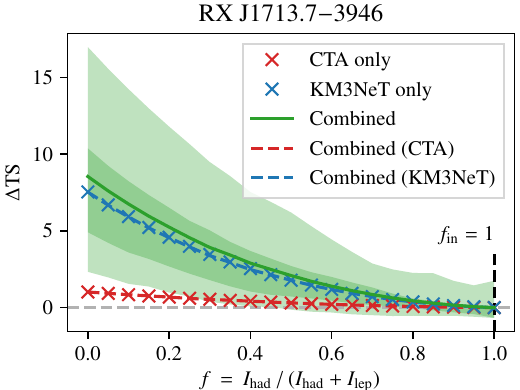}
  }
  \caption{
    Profile likelihood scan results for the sources \velax and \rxj.
    The plots on the left hand side display the results for a purely leptonic (IC) input model ($f_\mathrm{in}=0$), whereas the plots on the right hand side show those for a purely hadronic (PD) input model ($f_\mathrm{in}=1$).
    For more details, please refer to the caption of Fig.~\ref{fig:scan_velax}.
  }
  \label{fig:scans1}
\end{figure*}

\begin{figure*}
  \centering
  \subfigure{
    \includegraphics[width=0.48\textwidth]{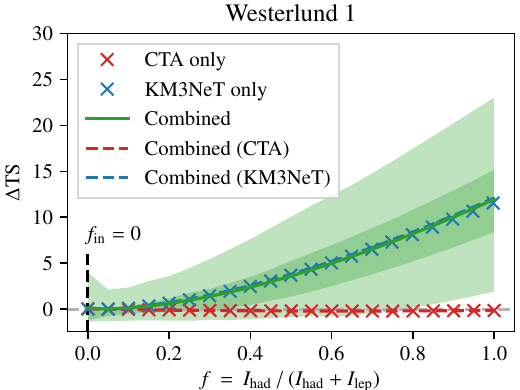}
  }
  \subfigure{
    \includegraphics[width=0.48\textwidth]{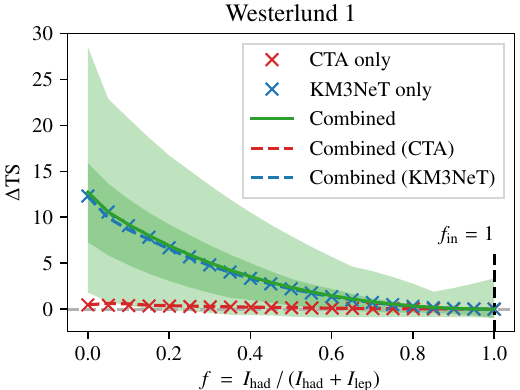}
  }\\
  \subfigure{
    \includegraphics[width=0.48\textwidth]{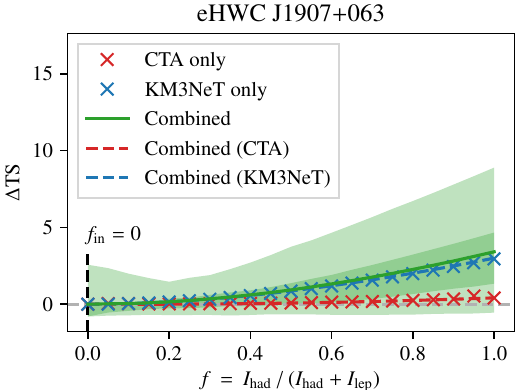}
  }
  \subfigure{
    \includegraphics[width=0.48\textwidth]{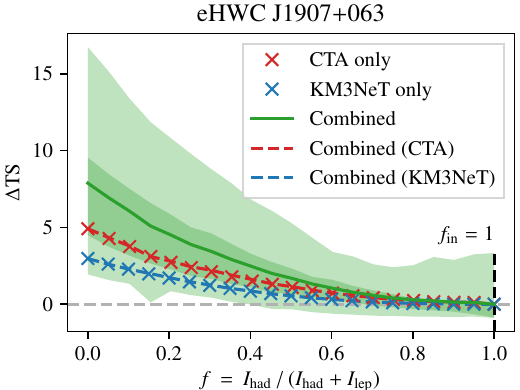}
  }
  \caption{Same as in Fig.~\ref{fig:scans1}, but for the sources \wld and \ehwc.}
  \label{fig:scans2}
\end{figure*}

Likelihood scan results for \velax and \rxj are shown in Fig.~\ref{fig:scans1}, while those for \wld and\\\ehwc are shown in Fig.~\ref{fig:scans2}.

\newpage

\section{Example of a PDF for the hadronic fraction $f$}
\label{sec:appendix_pdf}
\begin{figure}
  \centering
  \includegraphics{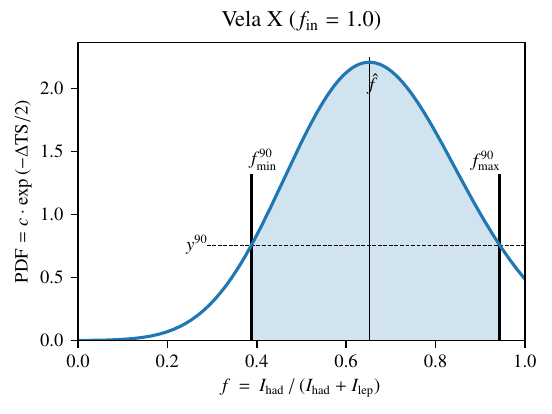}
  \caption{
    PDF for the hadronic fraction $f$ for a single pseudo experiment, constructed from the corresponding likelihood scan.
    The vertical line at the peak of the PDF marks the best-fit value $\hat{f}$, while the lines at $f_\mathrm{min}^{90}$ and  $f_\mathrm{max}^{90}$ show the lower and upper bound of the 90\% credible interval, respectively.
    The shaded area between the bounds contains 90\% of the total area between 0 and 1.
    The horizontal dashed line at $y^{90}$ shows that the intersections of the PDF with the bounds occur at the same height.
  }
  \label{fig:pdf}
\end{figure}

In Fig. \ref{fig:pdf}, we show an example of the PDF $\Phi(f)$ constructed from the likelihood scan (defined in Eq.~\ref{eq:pdf}) for an individual pseudo experiment in the hadronic scenario of \velax.
The pseudo experiment is hand-picked to show one of the few cases where the upper limit $f_\mathrm{max}^{90}$ does not coincide with 1.0.
We construct the limits such that all likelihood values included in the shaded interval are larger than those outside.
For this reason, the intersections of the limits with the PDF are at the same height $y^{90}$, which results in the smallest 90\% credible interval possible.

\end{appendix}

\bibliographystyle{spphys}
\bibliography{references}

\end{document}